\documentclass[%
 reprint,showkeys,
 amsmath,amssymb,
 aps,
]{revtex4-2}

\usepackage{subfig}
\usepackage{graphicx}
\usepackage{dcolumn}
\usepackage{bm}
\usepackage{hyperref}
\usepackage{float}
\usepackage{xcolor}

\newcommand{\be}{\begin{equation}}
\newcommand{\ee}{\end{equation}}
\newcommand{\lp}{\left(}
\newcommand{\rp}{\right)}

\DeclareMathOperator{\fvu}{\Gamma^{a,1}_{\mu}}
\DeclareMathOperator{\fvd}{\Gamma^{a,2}_{\mu}}
\DeclareMathOperator{\m}{\textit{m}_{\textit{f}}}
\DeclareMathOperator{\pu}{\textit{p}_1}
\DeclareMathOperator{\pd}{\textit{p}_2}

\DeclareMathOperator{\psu}{\not\!\textit{p}_1} 
 
\DeclareMathOperator{\psd}{\not\!\textit{p}_{2}}
\DeclareMathOperator{\psau}{\not\!\textit{p}_{1,\parallel}} 
\DeclareMathOperator{\psad}{\not\!\textit{p}_{2,\parallel}}

\DeclareMathOperator{\gm}{\gamma_{\mu}}
\DeclareMathOperator{\gma}{\gamma_{\mu}^{\parallel}}

\DeclareMathOperator{\ks}{\not\!\textit{k}} 
\DeclareMathOperator{\ka}{\textit{k}_{\parallel}}
\DeclareMathOperator{\ke}{\textit{k}_{\perp}} 

\begin{document}

\title{Magnetic corrections to the QCD coupling: strong field approximation}

\author{Gabriela Fernández}
\author{L. A. Hernández}%
\affiliation{Departamento de F\'isica, Universidad Aut\'onoma Metropolitana-Iztapalapa, Avenida San Rafael Atlixco 186, Ciudad de México 09340, Mexico.}%

\author{R. Zamora}
\affiliation{Instituto de Ciencias B\'asicas, Universidad Diego Portales, Casilla 298-V, Santiago, Chile.\\
Facultad de Medicina Veterinaria, Universidad San Sebasti\'an, Santiago, Chile.}

\date{\today}

\begin{abstract}
We compute the 1-loop vertex function of the QCD coupling in the presence of an ultra intense magnetic field. From the vertex function, we extract the effective coupling and show that it grows with increasing magnetic field. We consider the quark-gluon vertex and the three-gluon vertex, accounting for the propagators of charged particles within the loops using the lowest Landau level approximation in order to satisfy the condition where the magnetic field is the largest energy scale. Under this approximation, we find that the contribution from the three-gluon vertex vanishes. Therefore, this result  arises from the competition between the color charge associated to gluons and to quarks as well, with the former being larger than the latter. The behavior of the QCD coupling as a function of the magnetic field strength is analogous to that exhibited by the light-quark condensate, indicating the magnetic catalysis occurs. This increasing behavior stems from the dominant contribution of color charge associated to gluons in the vertex function.
\end{abstract}

\maketitle


In recent decades, there has been extensive study of the effects produced by strong magnetic fields on strongly interacting matter in extreme conditions~\cite{Kharzeev:2012ph}. This research is motivated by conditions found in systems such as Relativistic Heavy-Ion Collisions~\cite{Skokov:2009qp,Bzdak:2011yy,Voronyuk:2011jd,McLerran:2013hla,Brandenburg:2021lnj,STAR:2023jdd}, the cores of neutron stars~\cite{Ho:2011gy,Gusakov:2017uam,Igoshev:2021ewx} and the early universe~\cite{Grasso:2000wj,Subramanian:2009fu}. One of the most significant findings reported in the literature concerns the change in the \textit{vacuum expectation value} (vev) of the Dirac sector in the Quantum Chromodynamics,  referred to as the light-quark condensate, due to the presence of magnetic fields. When the magnetic field strength is large, the light-quark condensate varies, increasing as the magnetic field strength increases. This phenomenon is known as \textit{Magnetic Catalysis} (MC)~\cite{Gusynin:1995nb,Miransky:2002rp,Shovkovy:2012zn}. Another way to describe this phenomenon is to say that the breaking of chiral symmetry is reinforced thanks to the presence of a magnetic field. Results demonstrating MC from Lattice Quantum Chromodynamics (LQCD) were reported in Refs.~\cite{Bali:2012zg,Bali:2014kia,Kogut:2002zg,Bazavov:2017dus}. Additionally, there are results from effective models~\cite{Johnson:2008vna,Frasca:2011zn,Andersen:2021lnk,Bandyopadhyay:2020zte,G1} and holographic techniques~\cite{Zayakin:2008cy,Filev:2009xp,Ballon-Bayona:2020xtf} that provide explanations of the MC phenomenon. However, another question that arises is whether there are modifications in the coupling of strongly interacting fields when the system is permeated by a magnetic field. This idea can be addressed by working directly within the perturbative region of QCD, where we compute loop corrections order by order of the QCD vertices. The 1-loop correction of the QCD coupling at finite temperature and magnetic field, in the high temperature regime and weak field approximation, has been reported in Ref.~\cite{Ayala:2014uua}. In this study, the effective coupling constant exhibits a decreasing behavior as the magnetic field strength increases. It is important to highlight that this behavior is the same as that observed in the vacuum expectation value (vev) under the same conditions, suggesting that \textit{Inverse Magnetic Catalysis} occurs. In Ref.~\cite{Ayala:2015bgv} the correction to the quark-gluon vertex at zero temperature in the presence of a magnetic field was reported, focusing on the regime where this field is the smallest energy scale. The result showed a decreasing behavior of the effective coupling as the magnetic field increases. Once again, we observe that the effective coupling constant exhibits the same behavior as the vev. In these latter conditions, we observe Magnetic Catalysis.

To explore another condition where the QCD coupling may be affected by the presence of a magnetic field, we present in this work the computation of the QCD effective coupling at 1-loop order when a constant and uniform magnetic field permeates the system, assuming it to be the highest energy scale. In QCD, we have the quark-gluon vertex and pure gluonic vertices. The former is depicted in Fig.~\ref{fv1} (on the left), where a gluon with momentum $p_2-p_1$ enters to the vertex and a pair antiquark and quark leave the vertex, with momentum $p_1$ and $p_2$, respectively. 

\begin{figure}[b]
    \centering
    \includegraphics{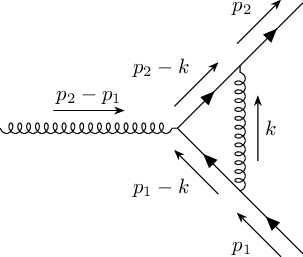}
    \caption{Feynman diagram of the QED-like contribution which includes the magnetic correction at 1-loop order to the quark-gluon vertex.}
    \label{fv2}
\end{figure}

\begin{figure}[b]
    \centering
    \includegraphics{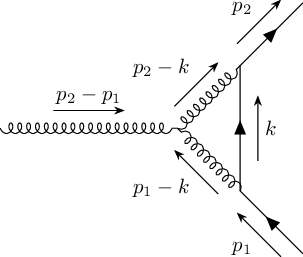}
    \caption{Feynman diagram of the pure QCD contribution which includes the magnetic correction at 1-loop order to the quark-gluon vertex.}
    \label{fv3}
\end{figure}

\begin{figure}[b]
    \centering
    \includegraphics{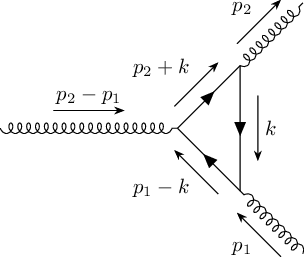}
    \caption{Feynman diagram which includes the magnetic correction at 1-loop order to the three-gluon vertex.}
    \label{fv4}
\end{figure}

In order to compute the 1-loop vertex function for this interaction, we consider two Feynman diagrams, depicted in Figs.~\ref{fv2} and~\ref{fv3}. The second type of vertex in QCD, depicted in Fig.~\ref{fv1} (on the right), involves pure gauge fields. In this case, there is one correction that could contribute to the 1-loop vertex function when a magnetic field is present. It corresponds to the case where three external gluons and three internal quarks generate the loop, as depicted in Fig.~\ref{fv4}. If the magnetic field is strong enough to be the largest energy scale, then we can work in the \textit{lowest Landau level} (LLL) approximation for the propagation of the charged fields. In this approximation, the loop in Fig.~\ref{fv4} is equal to zero, as shown in Refs.~\cite{Ayala:2017vex,Ayala:2022zhu}. Therefore, it is possible to work solely with the 1-loop correction to the quark-gluon vertex contribution.

\begin{widetext}
\center{
\begin{figure}[t]
\centering
    \subfloat[]{
           \includegraphics{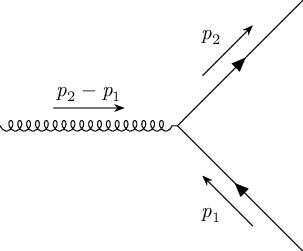}}
           \hspace{1.5cm}
    \subfloat[]{
        \includegraphics{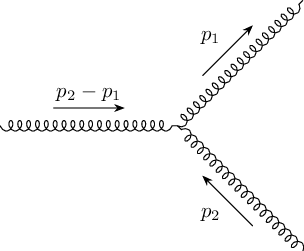}}
\caption{Feynman diagrams of the QCD interaction. The one on the left is the quark-gluon vertex and the one on the right is the three-gluon vertex}
\label{fv1} 
\end{figure}}
\end{widetext}

We start calculating the 1-loop vertex functions, where we split the computation in two terms. The first one involves a loop with two quarks and one gluon which identifies as $\Gamma_\mu^{a,1}$, while the second term involves a loop with two gluons and one quark which corresponds to $\Gamma_\mu^{a,2}$. The expressions for the vertex functions are
\begin{align}
i g\fvu =&\int \frac{d^4 k}{(2\pi)^4} i g t^b \gamma_{\nu} i S^{LLL}(\pd-k)igt^a \gm \nonumber\\
&\times i S^{LLL}(\pu-k) i g t^c \gamma_{\delta} i D_{\delta \nu}^{c b}(k), 
\label{vf1}
\end{align}
\begin{align}
ig\Gamma_{\mu}^{a,2}& = \int \frac{d^4k}{(2\pi)^4}i g t^b \gamma_{\nu} i S^{LLL}(k)ig t^{c_1}\gamma_{\alpha} \nonumber\\
& i D_{\alpha\beta}^{c_1 c_2}(p_2-k)i f^{a c_2c_3}i g \nonumber \\
&V_{\mu\beta\eta}(p_2-p_1,p_2-k,p_1-k) i D_{\eta \nu}^{c_3 b}(p_1-k),
\label{vf2}
\end{align}
where $igt^a \gamma_\mu $ is the quark-gluon vertex, and $if^{a_1 a_2 a_3}igV_{\mu_1 \mu_2 \mu_3 }(p_1,p_2,p_3)$ is the three-gluon vertex, with
\begin{align}
V_{\mu\beta\eta}&(p_2-p_1,p_2-k,p_1-k)=g_{\beta\eta}(p_1-p_2)_{\mu}\nonumber\\
&+(p_2-2p_1+k)^{\beta}g^{\eta\mu}+(p_1-k)^{\eta}g^{\mu\beta}.
\end{align}
The fermion propagator for a charged field in the presence of a magnetic field in the lowest Landau level approximation is written as follows
\begin{equation}
iS^{LLL}(k)=2i e^{-\frac{\ke^2}{|q_fB|}} \frac{\ks_{\parallel}+\m}{\ka^2-\m^2+i\epsilon}\mathcal{O}^+,
\label{fermionpropagator}
\end{equation} 
where the fermion's mass and the electric charge are $m_f$ and $q_f$, respectively. The projectors $\mathcal{O}^\pm$ are defined as
\begin{equation}
    \mathcal{O}^\pm \equiv \frac{1}{2} [1 \pm \gamma^1\gamma^2].
\end{equation} 
We are considering the magnetic field is taken as pointing along the $\hat{z}$-axis. As a consequence of the Lorentz symmetry breaking for charged particles, we rewrite the four vectors in two pieces, the parallel and the perpendicular components. Therefore, the four vectors can be written as follows 
\begin{equation}
   X^\mu X_\mu=(X_0^2-X_3^2)-(X_1^2+X_2^2)=X_\parallel^2-X_\perp^2.
   \label{components}
\end{equation}
The last element that remains to be defined in Eqs.~(\ref{vf1}) and (\ref{vf2}) is the gluon propagator, for this work, we choose the Feynman gauge and the propagator becomes
\begin{equation}
    iD^{ab}_{\mu\nu}(p)=\delta^{ab}\frac{-i g_{\mu\nu}}{p^2+i\epsilon}.
    \label{gaugebosonpropagator}
\end{equation}
Substituting Eqs.~(\ref{fermionpropagator}) and~(\ref{gaugebosonpropagator}) in Eqs.~(\ref{vf1}) and~(\ref{vf2}), and after some straightforward algebra within the numerator, the vertex corrections become  
\begin{align}
i g\fvu=&-8g^3 (C_F-\frac{C_A}{2}) t^a \int \frac{d^2k_{\perp}}{(2\pi)^2}\frac{d^2k_{\parallel}}{(2\pi)^2}\nonumber\\
&\times \frac{e^{-\frac{(\pd-k)^2_{\perp}}{|q_fB|}}e^{-\frac{(\pu-k)^2_{\perp}}{|q_fB|}}}{((\pd-k)_{\parallel}^2-\m^2)((\pu-k)_{\parallel}^2-\m^2)k^2}\nonumber\\
& \times \left( \gamma_{\mu}^{\parallel}\left( (\psd -\ks)_{\parallel} +\m \right) \left( (\psu -\ks)_{\parallel}+\m\right)  \right. \nonumber\\
&\left. -2\m (\pd+\pu-2k)_{\mu}^{\parallel}+\m^2 \gamma_{\mu}^{\parallel}\right),
\label{v1explicit}
\end{align}
\begin{align}
ig\Gamma_{\mu}^{a ,2} =&- C_A g^3 t^a \int \frac{d^2k_{\perp}}{(2\pi)^2}\frac{d^2k_{\parallel}}{(2\pi)^2}\nonumber\\
&\frac{e^{-\frac{k_{\perp}^2}{|q_fB|}}}{(k^2_{\parallel}-m^2)(p_2-k)^2(p_1-k)^2} \nonumber \\
&\times\left( (4m-2\ks_{\parallel})(\pu-\pd)_{\mu}+\gm (\ks_{\parallel}+m)\right.\nonumber\\
&\times \left. (\psd-2\psu+\ks)+(\psu-\ks)(\ks_{\parallel}+m)\gm \right),
\label{v2explicit}
\end{align}
where $C_A$ and $C_F$ are the color factors corresponding to the fundamental and adjoint representations of the $SU(N)$ Casimir operators. In order to integrate over all the momentum components, we implement the Feynman parametrization. Then we use the general formula
\begin{align}
    \frac{1}{A_1^{a_1}\cdots A_n^{a_n}}&=\frac{\Gamma[a_1+\cdots +a_n]}{\Gamma[a_1]\cdots \Gamma[a_n]}\int_0^\infty dx_1 \cdots \int_0^\infty dx_n \nonumber \\
    &\frac{\delta(1-\sum_{k=1}^nx_k)x_1^{a_1-1}\cdots x_n^{a_n-1}}{(\sum_{k=1}^n x_k A_k)^{\sum_{k=1}^n a_k}}.
    \label{feynmanparameters}
\end{align}

Therefore, we can rewrite the Eqs.~(\ref{v1explicit}) and~(\ref{v2explicit}), by using Eq.~(\ref{feynmanparameters}) with two Feynman parameters $x$ and $y$. Additionally, we manipulate the denominators  to express the parallel components of the integrals in terms of a pair of new variables, which are
\begin{align}
    \ell_{\parallel,1}&=k_{\parallel}-(y\pd+ y \pu), \nonumber \\
    \ell_{\parallel,2}&=k_{\parallel}-(y \pd+ (1-x-y) \pu).
    \label{parallelcomp12}
\end{align}
Then, we replace $k_\parallel$ in terms of $\ell_\parallel$ and observe that the integrals over each momentum component are within a symmetric range. Consequently, the integrals of odd powers of each parallel momentum component evaluate to zero. Thus, we obtain
\begin{align}
i g \fvu =& -16 g^3 \lp C_F-\frac{C_A}{2} \rp \int_0^1 dx\int_0^{1-x} dy \nonumber\\
&\int \frac{d^2 k_{\perp}}{(2\pi)^2}e^{-\frac{(\pd-k)_{\perp}^2}{|q_fB|}}e^{-\frac{(\pu-k)_{\perp}^2}{|q_fB|} }\nonumber\\
&\int \frac{d^2 \ell_{\parallel,1}}{(2\pi)^2}\frac{\gamma_{\alpha}^{\parallel}\gamma_{\mu}^{\parallel}\gamma_{\beta}^{\parallel}\ell_{\parallel}^{\alpha}\ell_{\parallel}^{\beta}+f_{\mu,1}^{\parallel}}{(\ell_{\parallel,1}^2-\Delta_1)^3},
\label{vertexfuncwithFeynman1}
\end{align}
\begin{align}
i g \fvd &=2 C_Ag^3 t^a \int_0^1 dx\int_0^{1-x} dy  \int \frac{d^2 k_{\perp}}{(2\pi)^2}e^{-\frac{k_{\perp}^2}{|q_fB|}} \nonumber\\
& \int \frac{d^2 \ell_{\parallel,2}}{(2\pi)^2}\frac{\ell_{\parallel}^{\alpha}\ell^{\nu}\gamma_{\alpha}^{\parallel}\gamma_{\nu}\gamma_{\mu}-\gamma_{\mu}\gamma_{\sigma}^{\parallel}\gamma_{\delta}\ell_{\parallel}^{\sigma}\ell^{\delta}-f_{\mu,2}}{(\ell_{\parallel,2}^2-\Delta_2)^3},
\label{vertexfuncwithFeynman2}
\end{align}
with
\begin{align}
\Delta_1=&(x p_{2,\parallel}+yp_{1,\parallel})^2+m^2(x+y)\nonumber\\
&+k_{\perp}^2(1-x-y)-x p_{2,\parallel}^2-yp_{1,\parallel}^2,
\label{expdelta1}
\end{align}
\begin{equation}
\begin{split}
\Delta_2=&(\left(\ke^2 (1-x)\right)+2 \ke p_{1,\perp} (x+y-1)\\
&-2 \ke p_{2,\perp} y+m^2 x+p_1^2 x+p_1^2 y-p_1^2\nonumber\\
&+(p_{2,\parallel} y-p_{1,\parallel} (x+y-1))^2-p_2^2 y),
\label{expdelta2}
\end{split}
\end{equation}
and
\begin{widetext}  
\begin{equation}
f_{\mu,1}^{\parallel}=\gma(\psd(1-x)-\psu y)(\psu(1-y)-\psd x)-2\m(\pd(1-2x)+\pu(1-2y))_{\mu}^{\parallel}+\m^2\gma),
\label{numeratorwithoutl1}
\end{equation}  
\begin{align}
f_{\mu,2}=&(4\m-2(\psau (1-x-y)+\psad y))(\pu-\pd)_{\mu}+\gamma_{\mu}(\psau(1-x-y)+\psad y)( \psad(1+y)\nonumber\\
&-\psau (1+x+y))
+(\psau(x+y)-\psad y)(\psau(1-x-y)+\psad y)\gamma_{\mu}.
\label{numeratorwithoutl2}
\end{align}
\end{widetext}

In order to compute each contribution, we proceed to carry out first the integral over the parallel components. Then the 1-loop corrections to the QCD vertex become
\begin{align}
ig\Gamma^{a,1}_{\mu} =&\frac{2i}{\pi}g^3\left( C_F-\frac{C_A}{2}\right) t^a\int_0^1dx\int_0^{1-x}dy\int\frac{d^2 k_{\perp}}{(2\pi)^2}\nonumber\\
&\times e^{-\frac{(p_2-k)_{\perp}^2}{|q_fB|}}e^{-\frac{(p_1-k)_{\perp}^2}{|q_fB|}} \lp
\frac{\gm^{\parallel}}{\Delta_1}+\frac{f_{\mu,1}^{\parallel}}{\Delta_1^2}\rp , \label{gamma1}
\end{align}
\begin{align}
ig\Gamma^{a,2}_{\mu}&=i \frac{C_A}{4\pi}g^3 t^{a}\int^1_0 dx \int_0^{1-x}dy \int \frac{d^2 \ke}{(2\pi)^2} e^{-\frac{\ke^2}{|q_fB|}}\frac{f_{\mu,2}}{\Delta_2^2}. \label{gamma2} 
\end{align}

Our next step is to perform the integral over the perpendicular components of momentum. However, before of integrating, we implement a suitable change of variable
\begin{align}
   Q=\pu-\pd,\\
   P=\frac{\pu+\pd}{2},
   \label{newmomenta}
\end{align}
the relative and the average momentum between the quark-pair, respectively. For simplicity we consider the symmetric three-momentum configuration, where $p_1=(E,\vec{p})$ and $-p_1=(E,-\vec{p})$. Thus, $Q=(2E,0)$ and $P=(0,\vec{p})$. It means that $Q^2$ is proportional to the energy and $P^2$ to the momentum squared carried by the gluon. Another important assumption is that we work in the static limit, $\vec{p}\rightarrow 0$, at the same time that we consider $Q^2$ large enough to guarantee that we are working in the perturbative region of QCD.

Since in Eqs.~(\ref{gamma1}) and~(\ref{gamma2}), the numerators do not depend on $k_\perp$, it is straightforward to identify two types of integrals to compute, which are
\begin{align}
\int d^2 k_{\perp} \frac{ e^{-\frac{2\ke^2}{|q_fB|}}}{a\ke^2+c}&=\frac{\pi e^{\frac{2 c}{a |q_fB|}} \Gamma \left(0,\frac{2 c}{a |q_fB|}\right)}{a} \nonumber \\
   &\equiv \frac{\pi  |q_fB|}{2 c } e^{\alpha} \Gamma \left(0,\alpha \right) \alpha,   \label{kpep1}
\end{align}
\begin{align}
\int d^2 k_{\perp} \frac{ e^{-\frac{2\ke^2}{|q_fB|}}}{(a\ke^2+c)^2}&=\frac{\pi}{a c}- \frac{2\pi e^{\frac{2 c}{a |q_fB|}} \Gamma \left(0,\frac{2 c}{a |q_fB|}\right)}{a^2 |q_fB|} \nonumber \\ 
&\equiv  \frac{\pi}{a c} - \frac{\pi}{a c} e^{\alpha} \Gamma \left(0,\alpha \right) \alpha,  \label{kpep2}
\end{align}
where
\begin{eqnarray}
 a&=&(1-x-y), \nonumber \\    
 c&=&\m^2 (x+y)+\frac{1}{4} Q^2 \left((x-y)^2-(x+y)\right), \nonumber \\   
 \alpha&=&\frac{2c}{a| q_fB|}.
\end{eqnarray}

In this step of the calculation process, it is relevant to remember that we are working in the large magnetic field limit, or in other words, we are working within the lowest Landau level approximation. Therefore, we are in the regime where $\alpha \ll 1$, and we get
\begin{equation}
\lim_{\alpha \rightarrow 0} e^{\alpha} \Gamma \left(0,\alpha \right) \alpha=1 + (1-\gamma_E) \alpha + \mathcal{O}(\alpha^2),
\label{aproximacion}
\end{equation}
where $\gamma_E$ is the Euler-Mascheroni constant. Then, we substitute Eq.~(\ref{aproximacion}) in Eqs.~(\ref{kpep1}) and~(\ref{kpep2}), and we obtain 
\begin{align}
ig\Gamma^{a,1}_{\mu} =&i g^3\left( C_F-\frac{C_A}{2}\right) t^a |q_fB| \gma \int_0^1dx\int_0^{1-x}dy \nonumber\\
&\times \frac{1}{m^2 (x+y)+\frac{1}{4} Q^2 \left((x-y)^2-(x+y)\right)},
\label{vertxfjustFeynman1}
\end{align}
\begin{align}
ig\Gamma^{a,2}_{\mu}&=0.
\label{vertxfjustFeynman2}
\end{align}

We proceed to integrate over the Feynman parameters, starting with the integral over $y$, we have
\begin{align}
ig&\Gamma^{a,1}_{\mu} =4i g^3\left( C_F-\frac{C_A}{2}\right) t^a |q_fB| \gma \int_0^1dx \nonumber\\
&\frac{1}{\sqrt{4\m^2-Q^2} \sqrt{4\m^2-Q^2 (8 x+1)}}\nonumber\\
&\left(\tanh ^{-1}\left(\frac{4\m^2-Q^2 (2 x+1)}{\sqrt{4\m^2-Q^2} \sqrt{4\m^2-Q^2 (8 x+1)}}\right)\nonumber \right.\\
&\left. -\tanh ^{-1}\left(\frac{4\m^2+Q^2 (1-4 x)}{\sqrt{4\m^2-Q^2} \sqrt{4\m^2-Q^2 (8 x+1)}}\right)\right).
\label{functionvertexwithoneparameter}
\end{align}
Before to perform the last integration, we rewrite the hyperbolic arctangent in terms of logarithms
\begin{widetext}
\begin{align}
   ig\Gamma^{a,1}_{\mu} =&4i g^3\left( C_F-\frac{C_A}{2}\right) t^a |q_fB| \gma \int_0^1dx \left(\ln \left(\frac{\sqrt{4\m^2-Q^2} \sqrt{4\m^2-Q^2 (8 x+1)}+4\m^2-Q^2 (2 x+1)}{\sqrt{4\m^2-Q^2} \sqrt{4\m^2-Q^2 (8 x+1)}-4\m^2+Q^2(2x+1) }\right)\right.\nonumber\\
   &+\left.\ln \left(\frac{\sqrt{4\m^2-Q^2} \sqrt{4\m^2-Q^2 (8 x+1)}-4\m^2+Q^2 (4 x-1)}{\sqrt{4\m^2-Q^2} \sqrt{4\m^2-Q^2 (8 x+1)}+4\m^2- Q^2(4 x-1)}\right)\right)\frac{1}{\sqrt{4\m^2-Q^2} \sqrt{4\m^2-Q^2 (8 x+1)}},
   \label{vertexfunctionlogarithms}
\end{align}
\end{widetext}
and we finally proceed to integrate the last Feynman parameter, $x$, being the final expression 
\begin{align}
 ig \fvu =&ig^3\left( C_F-\frac{C_A}{2}\right) t^a  \gma    \frac{|q_fB|}{Q^2}\left(\ln \left(1-\frac{Q^2}{4\m^2}\right)\right. \nonumber\\
 &\left.+\frac{2 Q \tan^{-1}\left(\frac{Q}{\sqrt{4\m^2-Q^2}}\right)}{\sqrt{4\m^2-Q^2}}\right).
 \label{finalexpressionvertexfunction}
\end{align}

From Eq.~(\ref{finalexpressionvertexfunction}), we can extract the effective QCD coupling in the presence of a very large magnetic field
\begin{align}
 g_{eff}&=g\left[ 1-g^2
 \frac{3|q_fB|}{2Q^2}\right. \nonumber \\
 &\left. \times\left(\ln \left(1-\frac{Q^2}{4\m^2}\right)+\frac{2 Q \tan^{-1}\left(\frac{Q}{\sqrt{4\m^2-Q^2}}\right)}{\sqrt{4\m^2-Q^2}}\right)\right],
 \label{geffective}
\end{align}
where we have used $C_F-\frac{C_A}{2}=-\frac{N_f}{2}$ and $N_f=3$.

Equation~(\ref{geffective}) represent our final result, displaying a monotonically increasing behavior of the coupling as the magnetic field strength increases. The crucial element in understanding the magnetic dependence of $g_{eff}$ is directly related to the coefficient $C_F-C_A/2$, which determines the behavior of this effective constant with changes in the magnetic field. Since the charge associated to the gluons is larger than the one associated to the quarks, we can conclude that the gluon dynamic is strengthened by an external magnetic field. It is noteworthy that this result holds even when the 1-loop order correction of the three-gluon vertex vanishes, a direct consequence of the lowest Landau level approximation. The Eq.~(\ref{geffective}) is consistent with $g_{eff}$ at $T=0$ and finite magnetic field, in the weak field limit, as shown in Ref.~\cite{Ayala:2015bgv}. Additionally, our result can be linked to the one reported in Ref.~\cite{Ayala:2014uua}, where the thermo-magnetic 1-loop correction to the quark-gluon vertex was computed in the limit of high temperature. In that case, the result is only proportional to $C_f$. Thus, it depends only on the color charge associated with the quarks, and the effective coupling has an opposite behavior. It decreases as the magnetic field strength increases.
\\

Notice that our analysis is within the perturbative regime of QCD. Hence, we consider $Q^2$, the gluon virtuality, to be large enough, at least larger than $1 \ GeV^2$, while also satisfying the relation $Q^2 < eB$, since we are operating under the strong magnetic field approximation. Concerning the kinematical conditions, we establish the configuration where the quark and antiquark travel back to back, implying that their relative orbital angular momentum $L$ vanishes. Since the gluon spin is equal to one, the pair quark-antiquark must carry a total spin $S=1$, aligned with the magnetic field. 
\\

For the situation when the magnetic field strength is ultra intense, we can utilize the result of this work to compute the $\bar{q}q$ scattering and annihilation processes. At first order in the perturbative series, not only should the correction of the field's propagation be relevant, but we can take into account that the strength of the interaction is enhanced by the presence of the magnetic field, providing a clear signal of the magnetic catalysis. However, there is still more work to be done. The non-perturbative region, where the gluon virtuality is not large enough, remains an open question. This is work currently under development and will be reported elsewhere.

\section*{Acknowledgements}

Support for this work was received in part by Consejo Nacional de Humanidades, Ciencia y Tecnolog\'ia grant number CF-2023-G-433. R. Zamora acknowledges support from FONDECYT (Chile) under grant No. 1241436. G. Fern\'andez acknowledge the financial support of a fellowship granted by Consejo Nacional de Humanidades, Ciencia y Tecnolog\'ia as part of the Sistema Nacional de Posgrados.

\bibliography{mybibliography}

\begin{thebibliography}{31}%
\makeatletter
\providecommand \@ifxundefined [1]{%
 \@ifx{#1\undefined}
}%
\providecommand \@ifnum [1]{%
 \ifnum #1\expandafter \@firstoftwo
 \else \expandafter \@secondoftwo
 \fi
}%
\providecommand \@ifx [1]{%
 \ifx #1\expandafter \@firstoftwo
 \else \expandafter \@secondoftwo
 \fi
}%
\providecommand \natexlab [1]{#1}%
\providecommand \enquote  [1]{``#1''}%
\providecommand \bibnamefont  [1]{#1}%
\providecommand \bibfnamefont [1]{#1}%
\providecommand \citenamefont [1]{#1}%
\providecommand \href@noop [0]{\@secondoftwo}%
\providecommand \href [0]{\begingroup \@sanitize@url \@href}%
\providecommand \@href[1]{\@@startlink{#1}\@@href}%
\providecommand \@@href[1]{\endgroup#1\@@endlink}%
\providecommand \@sanitize@url [0]{\catcode `\\12\catcode `\$12\catcode `\&12\catcode `\#12\catcode `\^12\catcode `\_12\catcode `\%12\relax}%
\providecommand \@@startlink[1]{}%
\providecommand \@@endlink[0]{}%
\providecommand \url  [0]{\begingroup\@sanitize@url \@url }%
\providecommand \@url [1]{\endgroup\@href {#1}{\urlprefix }}%
\providecommand \urlprefix  [0]{URL }%
\providecommand \Eprint [0]{\href }%
\providecommand \doibase [0]{https://doi.org/}%
\providecommand \selectlanguage [0]{\@gobble}%
\providecommand \bibinfo  [0]{\@secondoftwo}%
\providecommand \bibfield  [0]{\@secondoftwo}%
\providecommand \translation [1]{[#1]}%
\providecommand \BibitemOpen [0]{}%
\providecommand \bibitemStop [0]{}%
\providecommand \bibitemNoStop [0]{.\EOS\space}%
\providecommand \EOS [0]{\spacefactor3000\relax}%
\providecommand \BibitemShut  [1]{\csname bibitem#1\endcsname}%
\let\auto@bib@innerbib\@empty
\bibitem [{\citenamefont {Kharzeev}\ \emph {et~al.}(2013)\citenamefont {Kharzeev}, \citenamefont {Landsteiner}, \citenamefont {Schmitt},\ and\ \citenamefont {Yee}}]{Kharzeev:2012ph}%
  \BibitemOpen
  \bibfield  {author} {\bibinfo {author} {\bibfnamefont {D.~E.}\ \bibnamefont {Kharzeev}}, \bibinfo {author} {\bibfnamefont {K.}~\bibnamefont {Landsteiner}}, \bibinfo {author} {\bibfnamefont {A.}~\bibnamefont {Schmitt}},\ and\ \bibinfo {author} {\bibfnamefont {H.-U.}\ \bibnamefont {Yee}},\ }\bibfield  {title} {\bibinfo {title} {{'Strongly interacting matter in magnetic fields': an overview}},\ }\href {https://doi.org/10.1007/978-3-642-37305-3_1} {\bibfield  {journal} {\bibinfo  {journal} {Lect. Notes Phys.}\ }\textbf {\bibinfo {volume} {871}},\ \bibinfo {pages} {1} (\bibinfo {year} {2013})},\ \Eprint {https://arxiv.org/abs/1211.6245} {arXiv:1211.6245 [hep-ph]} \BibitemShut {NoStop}%
\bibitem [{\citenamefont {Skokov}\ \emph {et~al.}(2009)\citenamefont {Skokov}, \citenamefont {Illarionov},\ and\ \citenamefont {Toneev}}]{Skokov:2009qp}%
  \BibitemOpen
  \bibfield  {author} {\bibinfo {author} {\bibfnamefont {V.}~\bibnamefont {Skokov}}, \bibinfo {author} {\bibfnamefont {A.~Y.}\ \bibnamefont {Illarionov}},\ and\ \bibinfo {author} {\bibfnamefont {V.}~\bibnamefont {Toneev}},\ }\bibfield  {title} {\bibinfo {title} {{Estimate of the magnetic field strength in heavy-ion collisions}},\ }\href {https://doi.org/10.1142/S0217751X09047570} {\bibfield  {journal} {\bibinfo  {journal} {Int. J. Mod. Phys. A}\ }\textbf {\bibinfo {volume} {24}},\ \bibinfo {pages} {5925} (\bibinfo {year} {2009})},\ \Eprint {https://arxiv.org/abs/0907.1396} {arXiv:0907.1396 [nucl-th]} \BibitemShut {NoStop}%
\bibitem [{\citenamefont {Bzdak}\ and\ \citenamefont {Skokov}(2012)}]{Bzdak:2011yy}%
  \BibitemOpen
  \bibfield  {author} {\bibinfo {author} {\bibfnamefont {A.}~\bibnamefont {Bzdak}}\ and\ \bibinfo {author} {\bibfnamefont {V.}~\bibnamefont {Skokov}},\ }\bibfield  {title} {\bibinfo {title} {{Event-by-event fluctuations of magnetic and electric fields in heavy ion collisions}},\ }\href {https://doi.org/10.1016/j.physletb.2012.02.065} {\bibfield  {journal} {\bibinfo  {journal} {Phys. Lett. B}\ }\textbf {\bibinfo {volume} {710}},\ \bibinfo {pages} {171} (\bibinfo {year} {2012})},\ \Eprint {https://arxiv.org/abs/1111.1949} {arXiv:1111.1949 [hep-ph]} \BibitemShut {NoStop}%
\bibitem [{\citenamefont {Voronyuk}\ \emph {et~al.}(2011)\citenamefont {Voronyuk}, \citenamefont {Toneev}, \citenamefont {Cassing}, \citenamefont {Bratkovskaya}, \citenamefont {Konchakovski},\ and\ \citenamefont {Voloshin}}]{Voronyuk:2011jd}%
  \BibitemOpen
  \bibfield  {author} {\bibinfo {author} {\bibfnamefont {V.}~\bibnamefont {Voronyuk}}, \bibinfo {author} {\bibfnamefont {V.~D.}\ \bibnamefont {Toneev}}, \bibinfo {author} {\bibfnamefont {W.}~\bibnamefont {Cassing}}, \bibinfo {author} {\bibfnamefont {E.~L.}\ \bibnamefont {Bratkovskaya}}, \bibinfo {author} {\bibfnamefont {V.~P.}\ \bibnamefont {Konchakovski}},\ and\ \bibinfo {author} {\bibfnamefont {S.~A.}\ \bibnamefont {Voloshin}},\ }\bibfield  {title} {\bibinfo {title} {{(Electro-)Magnetic field evolution in relativistic heavy-ion collisions}},\ }\href {https://doi.org/10.1103/PhysRevC.83.054911} {\bibfield  {journal} {\bibinfo  {journal} {Phys. Rev. C}\ }\textbf {\bibinfo {volume} {83}},\ \bibinfo {pages} {054911} (\bibinfo {year} {2011})},\ \Eprint {https://arxiv.org/abs/1103.4239} {arXiv:1103.4239 [nucl-th]} \BibitemShut {NoStop}%
\bibitem [{\citenamefont {McLerran}\ and\ \citenamefont {Skokov}(2014)}]{McLerran:2013hla}%
  \BibitemOpen
  \bibfield  {author} {\bibinfo {author} {\bibfnamefont {L.}~\bibnamefont {McLerran}}\ and\ \bibinfo {author} {\bibfnamefont {V.}~\bibnamefont {Skokov}},\ }\bibfield  {title} {\bibinfo {title} {{Comments About the Electromagnetic Field in Heavy-Ion Collisions}},\ }\href {https://doi.org/10.1016/j.nuclphysa.2014.05.008} {\bibfield  {journal} {\bibinfo  {journal} {Nucl. Phys. A}\ }\textbf {\bibinfo {volume} {929}},\ \bibinfo {pages} {184} (\bibinfo {year} {2014})},\ \Eprint {https://arxiv.org/abs/1305.0774} {arXiv:1305.0774 [hep-ph]} \BibitemShut {NoStop}%
\bibitem [{\citenamefont {Brandenburg}\ \emph {et~al.}(2021)\citenamefont {Brandenburg}, \citenamefont {Zha},\ and\ \citenamefont {Xu}}]{Brandenburg:2021lnj}%
  \BibitemOpen
  \bibfield  {author} {\bibinfo {author} {\bibfnamefont {J.~D.}\ \bibnamefont {Brandenburg}}, \bibinfo {author} {\bibfnamefont {W.}~\bibnamefont {Zha}},\ and\ \bibinfo {author} {\bibfnamefont {Z.}~\bibnamefont {Xu}},\ }\bibfield  {title} {\bibinfo {title} {{Mapping the electromagnetic fields of heavy-ion collisions with the Breit-Wheeler process}},\ }\href {https://doi.org/10.1140/epja/s10050-021-00595-5} {\bibfield  {journal} {\bibinfo  {journal} {Eur. Phys. J. A}\ }\textbf {\bibinfo {volume} {57}},\ \bibinfo {pages} {299} (\bibinfo {year} {2021})},\ \Eprint {https://arxiv.org/abs/2103.16623} {arXiv:2103.16623 [hep-ph]} \BibitemShut {NoStop}%
\bibitem [{\citenamefont {Abdulhamid}\ \emph {et~al.}(2024)\citenamefont {Abdulhamid} \emph {et~al.}}]{STAR:2023jdd}%
  \BibitemOpen
  \bibfield  {author} {\bibinfo {author} {\bibfnamefont {M.~I.}\ \bibnamefont {Abdulhamid}} \emph {et~al.} (\bibinfo {collaboration} {STAR}),\ }\bibfield  {title} {\bibinfo {title} {{Observation of the electromagnetic field effect via charge-dependent directed flow in heavy-ion collisions at the Relativistic Heavy Ion Collider}},\ }\href {https://doi.org/10.1103/PhysRevX.14.011028} {\bibfield  {journal} {\bibinfo  {journal} {Phys. Rev. X}\ }\textbf {\bibinfo {volume} {14}},\ \bibinfo {pages} {011028} (\bibinfo {year} {2024})},\ \Eprint {https://arxiv.org/abs/2304.03430} {arXiv:2304.03430 [nucl-ex]} \BibitemShut {NoStop}%
\bibitem [{\citenamefont {Ho}(2011)}]{Ho:2011gy}%
  \BibitemOpen
  \bibfield  {author} {\bibinfo {author} {\bibfnamefont {W.~C.~G.}\ \bibnamefont {Ho}},\ }\bibfield  {title} {\bibinfo {title} {{Evolution of a buried magnetic field in the central compact object neutron stars}},\ }\href {https://doi.org/10.1111/j.1365-2966.2011.18576.x} {\bibfield  {journal} {\bibinfo  {journal} {Mon. Not. Roy. Astron. Soc.}\ }\textbf {\bibinfo {volume} {414}},\ \bibinfo {pages} {2567} (\bibinfo {year} {2011})},\ \Eprint {https://arxiv.org/abs/1102.4870} {arXiv:1102.4870 [astro-ph.HE]} \BibitemShut {NoStop}%
\bibitem [{\citenamefont {Gusakov}\ \emph {et~al.}(2017)\citenamefont {Gusakov}, \citenamefont {Kantor},\ and\ \citenamefont {Ofengeim}}]{Gusakov:2017uam}%
  \BibitemOpen
  \bibfield  {author} {\bibinfo {author} {\bibfnamefont {M.~E.}\ \bibnamefont {Gusakov}}, \bibinfo {author} {\bibfnamefont {E.~M.}\ \bibnamefont {Kantor}},\ and\ \bibinfo {author} {\bibfnamefont {D.~D.}\ \bibnamefont {Ofengeim}},\ }\bibfield  {title} {\bibinfo {title} {{On the evolution of magnetic field in neutron stars}},\ }\href {https://doi.org/10.1103/PhysRevD.96.103012} {\bibfield  {journal} {\bibinfo  {journal} {Phys. Rev. D}\ }\textbf {\bibinfo {volume} {96}},\ \bibinfo {pages} {103012} (\bibinfo {year} {2017})},\ \Eprint {https://arxiv.org/abs/1705.00508} {arXiv:1705.00508 [astro-ph.HE]} \BibitemShut {NoStop}%
\bibitem [{\citenamefont {Igoshev}\ \emph {et~al.}(2021)\citenamefont {Igoshev}, \citenamefont {Popov},\ and\ \citenamefont {Hollerbach}}]{Igoshev:2021ewx}%
  \BibitemOpen
  \bibfield  {author} {\bibinfo {author} {\bibfnamefont {A.~P.}\ \bibnamefont {Igoshev}}, \bibinfo {author} {\bibfnamefont {S.~B.}\ \bibnamefont {Popov}},\ and\ \bibinfo {author} {\bibfnamefont {R.}~\bibnamefont {Hollerbach}},\ }\bibfield  {title} {\bibinfo {title} {{Evolution of Neutron Star Magnetic Fields}},\ }\href {https://doi.org/10.3390/universe7090351} {\bibfield  {journal} {\bibinfo  {journal} {Universe}\ }\textbf {\bibinfo {volume} {7}},\ \bibinfo {pages} {351} (\bibinfo {year} {2021})},\ \Eprint {https://arxiv.org/abs/2109.05584} {arXiv:2109.05584 [astro-ph.HE]} \BibitemShut {NoStop}%
\bibitem [{\citenamefont {Grasso}\ and\ \citenamefont {Rubinstein}(2001)}]{Grasso:2000wj}%
  \BibitemOpen
  \bibfield  {author} {\bibinfo {author} {\bibfnamefont {D.}~\bibnamefont {Grasso}}\ and\ \bibinfo {author} {\bibfnamefont {H.~R.}\ \bibnamefont {Rubinstein}},\ }\bibfield  {title} {\bibinfo {title} {{Magnetic fields in the early universe}},\ }\href {https://doi.org/10.1016/S0370-1573(00)00110-1} {\bibfield  {journal} {\bibinfo  {journal} {Phys. Rept.}\ }\textbf {\bibinfo {volume} {348}},\ \bibinfo {pages} {163} (\bibinfo {year} {2001})},\ \Eprint {https://arxiv.org/abs/astro-ph/0009061} {arXiv:astro-ph/0009061} \BibitemShut {NoStop}%
\bibitem [{\citenamefont {Subramanian}(2010)}]{Subramanian:2009fu}%
  \BibitemOpen
  \bibfield  {author} {\bibinfo {author} {\bibfnamefont {K.}~\bibnamefont {Subramanian}},\ }\bibfield  {title} {\bibinfo {title} {{Magnetic fields in the early universe}},\ }\href {https://doi.org/10.1002/asna.200911312} {\bibfield  {journal} {\bibinfo  {journal} {Astron. Nachr.}\ }\textbf {\bibinfo {volume} {331}},\ \bibinfo {pages} {110} (\bibinfo {year} {2010})},\ \Eprint {https://arxiv.org/abs/0911.4771} {arXiv:0911.4771 [astro-ph.CO]} \BibitemShut {NoStop}%
\bibitem [{\citenamefont {Gusynin}\ \emph {et~al.}(1996)\citenamefont {Gusynin}, \citenamefont {Miransky},\ and\ \citenamefont {Shovkovy}}]{Gusynin:1995nb}%
  \BibitemOpen
  \bibfield  {author} {\bibinfo {author} {\bibfnamefont {V.~P.}\ \bibnamefont {Gusynin}}, \bibinfo {author} {\bibfnamefont {V.~A.}\ \bibnamefont {Miransky}},\ and\ \bibinfo {author} {\bibfnamefont {I.~A.}\ \bibnamefont {Shovkovy}},\ }\bibfield  {title} {\bibinfo {title} {{Dimensional reduction and catalysis of dynamical symmetry breaking by a magnetic field}},\ }\href {https://doi.org/10.1016/0550-3213(96)00021-1} {\bibfield  {journal} {\bibinfo  {journal} {Nucl. Phys. B}\ }\textbf {\bibinfo {volume} {462}},\ \bibinfo {pages} {249} (\bibinfo {year} {1996})},\ \Eprint {https://arxiv.org/abs/hep-ph/9509320} {arXiv:hep-ph/9509320} \BibitemShut {NoStop}%
\bibitem [{\citenamefont {Miransky}\ and\ \citenamefont {Shovkovy}(2002)}]{Miransky:2002rp}%
  \BibitemOpen
  \bibfield  {author} {\bibinfo {author} {\bibfnamefont {V.~A.}\ \bibnamefont {Miransky}}\ and\ \bibinfo {author} {\bibfnamefont {I.~A.}\ \bibnamefont {Shovkovy}},\ }\bibfield  {title} {\bibinfo {title} {{Magnetic catalysis and anisotropic confinement in QCD}},\ }\href {https://doi.org/10.1103/PhysRevD.66.045006} {\bibfield  {journal} {\bibinfo  {journal} {Phys. Rev. D}\ }\textbf {\bibinfo {volume} {66}},\ \bibinfo {pages} {045006} (\bibinfo {year} {2002})},\ \Eprint {https://arxiv.org/abs/hep-ph/0205348} {arXiv:hep-ph/0205348} \BibitemShut {NoStop}%
\bibitem [{\citenamefont {Shovkovy}(2013)}]{Shovkovy:2012zn}%
  \BibitemOpen
  \bibfield  {author} {\bibinfo {author} {\bibfnamefont {I.~A.}\ \bibnamefont {Shovkovy}},\ }\bibfield  {title} {\bibinfo {title} {{Magnetic Catalysis: A Review}},\ }\href {https://doi.org/10.1007/978-3-642-37305-3_2} {\bibfield  {journal} {\bibinfo  {journal} {Lect. Notes Phys.}\ }\textbf {\bibinfo {volume} {871}},\ \bibinfo {pages} {13} (\bibinfo {year} {2013})},\ \Eprint {https://arxiv.org/abs/1207.5081} {arXiv:1207.5081 [hep-ph]} \BibitemShut {NoStop}%
\bibitem [{\citenamefont {Bali}\ \emph {et~al.}(2012)\citenamefont {Bali}, \citenamefont {Bruckmann}, \citenamefont {Endrodi}, \citenamefont {Fodor}, \citenamefont {Katz},\ and\ \citenamefont {Schafer}}]{Bali:2012zg}%
  \BibitemOpen
  \bibfield  {author} {\bibinfo {author} {\bibfnamefont {G.~S.}\ \bibnamefont {Bali}}, \bibinfo {author} {\bibfnamefont {F.}~\bibnamefont {Bruckmann}}, \bibinfo {author} {\bibfnamefont {G.}~\bibnamefont {Endrodi}}, \bibinfo {author} {\bibfnamefont {Z.}~\bibnamefont {Fodor}}, \bibinfo {author} {\bibfnamefont {S.~D.}\ \bibnamefont {Katz}},\ and\ \bibinfo {author} {\bibfnamefont {A.}~\bibnamefont {Schafer}},\ }\bibfield  {title} {\bibinfo {title} {{QCD quark condensate in external magnetic fields}},\ }\href {https://doi.org/10.1103/PhysRevD.86.071502} {\bibfield  {journal} {\bibinfo  {journal} {Phys. Rev. D}\ }\textbf {\bibinfo {volume} {86}},\ \bibinfo {pages} {071502} (\bibinfo {year} {2012})},\ \Eprint {https://arxiv.org/abs/1206.4205} {arXiv:1206.4205 [hep-lat]} \BibitemShut {NoStop}%
\bibitem [{\citenamefont {Bali}\ \emph {et~al.}(2014)\citenamefont {Bali}, \citenamefont {Bruckmann}, \citenamefont {Endr\"odi}, \citenamefont {Katz},\ and\ \citenamefont {Sch\"afer}}]{Bali:2014kia}%
  \BibitemOpen
  \bibfield  {author} {\bibinfo {author} {\bibfnamefont {G.~S.}\ \bibnamefont {Bali}}, \bibinfo {author} {\bibfnamefont {F.}~\bibnamefont {Bruckmann}}, \bibinfo {author} {\bibfnamefont {G.}~\bibnamefont {Endr\"odi}}, \bibinfo {author} {\bibfnamefont {S.~D.}\ \bibnamefont {Katz}},\ and\ \bibinfo {author} {\bibfnamefont {A.}~\bibnamefont {Sch\"afer}},\ }\bibfield  {title} {\bibinfo {title} {{The QCD equation of state in background magnetic fields}},\ }\href {https://doi.org/10.1007/JHEP08(2014)177} {\bibfield  {journal} {\bibinfo  {journal} {JHEP}\ }\textbf {\bibinfo {volume} {08}},\ \bibinfo {pages} {177}},\ \Eprint {https://arxiv.org/abs/1406.0269} {arXiv:1406.0269 [hep-lat]} \BibitemShut {NoStop}%
\bibitem [{\citenamefont {Kogut}\ and\ \citenamefont {Sinclair}(2002)}]{Kogut:2002zg}%
  \BibitemOpen
  \bibfield  {author} {\bibinfo {author} {\bibfnamefont {J.~B.}\ \bibnamefont {Kogut}}\ and\ \bibinfo {author} {\bibfnamefont {D.~K.}\ \bibnamefont {Sinclair}},\ }\bibfield  {title} {\bibinfo {title} {{Lattice QCD at finite isospin density at zero and finite temperature}},\ }\href {https://doi.org/10.1103/PhysRevD.66.034505} {\bibfield  {journal} {\bibinfo  {journal} {Phys. Rev. D}\ }\textbf {\bibinfo {volume} {66}},\ \bibinfo {pages} {034505} (\bibinfo {year} {2002})},\ \Eprint {https://arxiv.org/abs/hep-lat/0202028} {arXiv:hep-lat/0202028} \BibitemShut {NoStop}%
\bibitem [{\citenamefont {Bazavov}\ \emph {et~al.}(2017)\citenamefont {Bazavov} \emph {et~al.}}]{Bazavov:2017dus}%
  \BibitemOpen
  \bibfield  {author} {\bibinfo {author} {\bibfnamefont {A.}~\bibnamefont {Bazavov}} \emph {et~al.},\ }\bibfield  {title} {\bibinfo {title} {{The QCD Equation of State to $\mathcal{O}(\mu_B^6)$ from Lattice QCD}},\ }\href {https://doi.org/10.1103/PhysRevD.95.054504} {\bibfield  {journal} {\bibinfo  {journal} {Phys. Rev. D}\ }\textbf {\bibinfo {volume} {95}},\ \bibinfo {pages} {054504} (\bibinfo {year} {2017})},\ \Eprint {https://arxiv.org/abs/1701.04325} {arXiv:1701.04325 [hep-lat]} \BibitemShut {NoStop}%
\bibitem [{\citenamefont {Johnson}\ and\ \citenamefont {Kundu}(2008)}]{Johnson:2008vna}%
  \BibitemOpen
  \bibfield  {author} {\bibinfo {author} {\bibfnamefont {C.~V.}\ \bibnamefont {Johnson}}\ and\ \bibinfo {author} {\bibfnamefont {A.}~\bibnamefont {Kundu}},\ }\bibfield  {title} {\bibinfo {title} {{External Fields and Chiral Symmetry Breaking in the Sakai-Sugimoto Model}},\ }\href {https://doi.org/10.1088/1126-6708/2008/12/053} {\bibfield  {journal} {\bibinfo  {journal} {JHEP}\ }\textbf {\bibinfo {volume} {12}},\ \bibinfo {pages} {053}},\ \Eprint {https://arxiv.org/abs/0803.0038} {arXiv:0803.0038 [hep-th]} \BibitemShut {NoStop}%
\bibitem [{\citenamefont {Frasca}\ and\ \citenamefont {Ruggieri}(2011)}]{Frasca:2011zn}%
  \BibitemOpen
  \bibfield  {author} {\bibinfo {author} {\bibfnamefont {M.}~\bibnamefont {Frasca}}\ and\ \bibinfo {author} {\bibfnamefont {M.}~\bibnamefont {Ruggieri}},\ }\bibfield  {title} {\bibinfo {title} {{Magnetic Susceptibility of the Quark Condensate and Polarization from Chiral Models}},\ }\href {https://doi.org/10.1103/PhysRevD.83.094024} {\bibfield  {journal} {\bibinfo  {journal} {Phys. Rev. D}\ }\textbf {\bibinfo {volume} {83}},\ \bibinfo {pages} {094024} (\bibinfo {year} {2011})},\ \Eprint {https://arxiv.org/abs/1103.1194} {arXiv:1103.1194 [hep-ph]} \BibitemShut {NoStop}%
\bibitem [{\citenamefont {Andersen}(2021)}]{Andersen:2021lnk}%
  \BibitemOpen
  \bibfield  {author} {\bibinfo {author} {\bibfnamefont {J.~O.}\ \bibnamefont {Andersen}},\ }\bibfield  {title} {\bibinfo {title} {{QCD phase diagram in a constant magnetic background: Inverse magnetic catalysis: where models meet the lattice}},\ }\href {https://doi.org/10.1140/epja/s10050-021-00491-y} {\bibfield  {journal} {\bibinfo  {journal} {Eur. Phys. J. A}\ }\textbf {\bibinfo {volume} {57}},\ \bibinfo {pages} {189} (\bibinfo {year} {2021})},\ \Eprint {https://arxiv.org/abs/2102.13165} {arXiv:2102.13165 [hep-ph]} \BibitemShut {NoStop}%
\bibitem [{\citenamefont {Bandyopadhyay}\ and\ \citenamefont {Farias}(2021)}]{Bandyopadhyay:2020zte}%
  \BibitemOpen
  \bibfield  {author} {\bibinfo {author} {\bibfnamefont {A.}~\bibnamefont {Bandyopadhyay}}\ and\ \bibinfo {author} {\bibfnamefont {R.~L.~S.}\ \bibnamefont {Farias}},\ }\bibfield  {title} {\bibinfo {title} {{Inverse magnetic catalysis: how much do we know about?}},\ }\href {https://doi.org/10.1140/epjs/s11734-021-00023-1} {\bibfield  {journal} {\bibinfo  {journal} {Eur. Phys. J. ST}\ }\textbf {\bibinfo {volume} {230}},\ \bibinfo {pages} {719} (\bibinfo {year} {2021})},\ \Eprint {https://arxiv.org/abs/2003.11054} {arXiv:2003.11054 [hep-ph]} \BibitemShut {NoStop}%
\bibitem [{\citenamefont {Gaspar}\ \emph {et~al.}(2023)\citenamefont {Gaspar}, \citenamefont {Hern\'andez},\ and\ \citenamefont {Zamora}}]{G1}%
  \BibitemOpen
  \bibfield  {author} {\bibinfo {author} {\bibfnamefont {I.~I.}\ \bibnamefont {Gaspar}}, \bibinfo {author} {\bibfnamefont {L.~A.}\ \bibnamefont {Hern\'andez}},\ and\ \bibinfo {author} {\bibfnamefont {R.}~\bibnamefont {Zamora}},\ }\bibfield  {title} {\bibinfo {title} {{Chiral symmetry restoration in a rotating medium}},\ }\href {https://doi.org/10.1103/PhysRevD.108.094020} {\bibfield  {journal} {\bibinfo  {journal} {Phys. Rev. D}\ }\textbf {\bibinfo {volume} {108}},\ \bibinfo {pages} {094020} (\bibinfo {year} {2023})},\ \Eprint {https://arxiv.org/abs/2305.00101} {arXiv:2305.00101 [hep-ph]} \BibitemShut {NoStop}%
\bibitem [{\citenamefont {Zayakin}(2008)}]{Zayakin:2008cy}%
  \BibitemOpen
  \bibfield  {author} {\bibinfo {author} {\bibfnamefont {A.~V.}\ \bibnamefont {Zayakin}},\ }\bibfield  {title} {\bibinfo {title} {{QCD Vacuum Properties in a Magnetic Field from AdS/CFT: Chiral Condensate and Goldstone Mass}},\ }\href {https://doi.org/10.1088/1126-6708/2008/07/116} {\bibfield  {journal} {\bibinfo  {journal} {JHEP}\ }\textbf {\bibinfo {volume} {07}},\ \bibinfo {pages} {116}},\ \Eprint {https://arxiv.org/abs/0807.2917} {arXiv:0807.2917 [hep-th]} \BibitemShut {NoStop}%
\bibitem [{\citenamefont {Filev}\ \emph {et~al.}(2009)\citenamefont {Filev}, \citenamefont {Johnson},\ and\ \citenamefont {Shock}}]{Filev:2009xp}%
  \BibitemOpen
  \bibfield  {author} {\bibinfo {author} {\bibfnamefont {V.~G.}\ \bibnamefont {Filev}}, \bibinfo {author} {\bibfnamefont {C.~V.}\ \bibnamefont {Johnson}},\ and\ \bibinfo {author} {\bibfnamefont {J.~P.}\ \bibnamefont {Shock}},\ }\bibfield  {title} {\bibinfo {title} {{Universal Holographic Chiral Dynamics in an External Magnetic Field}},\ }\href {https://doi.org/10.1088/1126-6708/2009/08/013} {\bibfield  {journal} {\bibinfo  {journal} {JHEP}\ }\textbf {\bibinfo {volume} {08}},\ \bibinfo {pages} {013}},\ \Eprint {https://arxiv.org/abs/0903.5345} {arXiv:0903.5345 [hep-th]} \BibitemShut {NoStop}%
\bibitem [{\citenamefont {Ballon-Bayona}\ \emph {et~al.}(2020)\citenamefont {Ballon-Bayona}, \citenamefont {Shock},\ and\ \citenamefont {Zoakos}}]{Ballon-Bayona:2020xtf}%
  \BibitemOpen
  \bibfield  {author} {\bibinfo {author} {\bibfnamefont {A.}~\bibnamefont {Ballon-Bayona}}, \bibinfo {author} {\bibfnamefont {J.~P.}\ \bibnamefont {Shock}},\ and\ \bibinfo {author} {\bibfnamefont {D.}~\bibnamefont {Zoakos}},\ }\bibfield  {title} {\bibinfo {title} {{Magnetic catalysis and the chiral condensate in holographic QCD}},\ }\href {https://doi.org/10.1007/JHEP10(2020)193} {\bibfield  {journal} {\bibinfo  {journal} {JHEP}\ }\textbf {\bibinfo {volume} {10}},\ \bibinfo {pages} {193}},\ \Eprint {https://arxiv.org/abs/2005.00500} {arXiv:2005.00500 [hep-th]} \BibitemShut {NoStop}%
\bibitem [{\citenamefont {Ayala}\ \emph {et~al.}(2015)\citenamefont {Ayala}, \citenamefont {Cobos-Mart\'\i{}nez}, \citenamefont {Loewe}, \citenamefont {Tejeda-Yeomans},\ and\ \citenamefont {Zamora}}]{Ayala:2014uua}%
  \BibitemOpen
  \bibfield  {author} {\bibinfo {author} {\bibfnamefont {A.}~\bibnamefont {Ayala}}, \bibinfo {author} {\bibfnamefont {J.~J.}\ \bibnamefont {Cobos-Mart\'\i{}nez}}, \bibinfo {author} {\bibfnamefont {M.}~\bibnamefont {Loewe}}, \bibinfo {author} {\bibfnamefont {M.~E.}\ \bibnamefont {Tejeda-Yeomans}},\ and\ \bibinfo {author} {\bibfnamefont {R.}~\bibnamefont {Zamora}},\ }\bibfield  {title} {\bibinfo {title} {{Finite temperature quark-gluon vertex with a magnetic field in the Hard Thermal Loop approximation}},\ }\href {https://doi.org/10.1103/PhysRevD.91.016007} {\bibfield  {journal} {\bibinfo  {journal} {Phys. Rev. D}\ }\textbf {\bibinfo {volume} {91}},\ \bibinfo {pages} {016007} (\bibinfo {year} {2015})},\ \Eprint {https://arxiv.org/abs/1410.6388} {arXiv:1410.6388 [hep-ph]} \BibitemShut {NoStop}%
\bibitem [{\citenamefont {Ayala}\ \emph {et~al.}(2016)\citenamefont {Ayala}, \citenamefont {Dominguez}, \citenamefont {Hernandez}, \citenamefont {Loewe},\ and\ \citenamefont {Zamora}}]{Ayala:2015bgv}%
  \BibitemOpen
  \bibfield  {author} {\bibinfo {author} {\bibfnamefont {A.}~\bibnamefont {Ayala}}, \bibinfo {author} {\bibfnamefont {C.~A.}\ \bibnamefont {Dominguez}}, \bibinfo {author} {\bibfnamefont {L.~A.}\ \bibnamefont {Hernandez}}, \bibinfo {author} {\bibfnamefont {M.}~\bibnamefont {Loewe}},\ and\ \bibinfo {author} {\bibfnamefont {R.}~\bibnamefont {Zamora}},\ }\bibfield  {title} {\bibinfo {title} {{Inverse magnetic catalysis from the properties of the QCD coupling in a magnetic field}},\ }\href {https://doi.org/10.1016/j.physletb.2016.05.058} {\bibfield  {journal} {\bibinfo  {journal} {Phys. Lett. B}\ }\textbf {\bibinfo {volume} {759}},\ \bibinfo {pages} {99} (\bibinfo {year} {2016})},\ \Eprint {https://arxiv.org/abs/1510.09134} {arXiv:1510.09134 [hep-ph]} \BibitemShut {NoStop}%
\bibitem [{\citenamefont {Ayala}\ \emph {et~al.}(2017)\citenamefont {Ayala}, \citenamefont {Castano-Yepes}, \citenamefont {Dominguez}, \citenamefont {Hernandez}, \citenamefont {Hernandez-Ortiz},\ and\ \citenamefont {Tejeda-Yeomans}}]{Ayala:2017vex}%
  \BibitemOpen
  \bibfield  {author} {\bibinfo {author} {\bibfnamefont {A.}~\bibnamefont {Ayala}}, \bibinfo {author} {\bibfnamefont {J.~D.}\ \bibnamefont {Castano-Yepes}}, \bibinfo {author} {\bibfnamefont {C.~A.}\ \bibnamefont {Dominguez}}, \bibinfo {author} {\bibfnamefont {L.~A.}\ \bibnamefont {Hernandez}}, \bibinfo {author} {\bibfnamefont {S.}~\bibnamefont {Hernandez-Ortiz}},\ and\ \bibinfo {author} {\bibfnamefont {M.~E.}\ \bibnamefont {Tejeda-Yeomans}},\ }\bibfield  {title} {\bibinfo {title} {{Prompt photon yield and elliptic flow from gluon fusion induced by magnetic fields in relativistic heavy-ion collisions}},\ }\href {https://doi.org/10.1103/PhysRevD.96.014023} {\bibfield  {journal} {\bibinfo  {journal} {Phys. Rev. D}\ }\textbf {\bibinfo {volume} {96}},\ \bibinfo {pages} {014023} (\bibinfo {year} {2017})},\ \bibinfo {note} {[Erratum: Phys.Rev.D 96, 119901 (2017)]},\ \Eprint {https://arxiv.org/abs/1704.02433} {arXiv:1704.02433 [hep-ph]} \BibitemShut {NoStop}%
\bibitem [{\citenamefont {Ayala}\ \emph {et~al.}(2022)\citenamefont {Ayala}, \citenamefont {Casta\~no Yepes}, \citenamefont {Hern\'andez}, \citenamefont {Mizher}, \citenamefont {Tejeda-Yeomans},\ and\ \citenamefont {Zamora}}]{Ayala:2022zhu}%
  \BibitemOpen
  \bibfield  {author} {\bibinfo {author} {\bibfnamefont {A.}~\bibnamefont {Ayala}}, \bibinfo {author} {\bibfnamefont {J.~D.}\ \bibnamefont {Casta\~no Yepes}}, \bibinfo {author} {\bibfnamefont {L.~A.}\ \bibnamefont {Hern\'andez}}, \bibinfo {author} {\bibfnamefont {A.~J.}\ \bibnamefont {Mizher}}, \bibinfo {author} {\bibfnamefont {M.~E.}\ \bibnamefont {Tejeda-Yeomans}},\ and\ \bibinfo {author} {\bibfnamefont {R.}~\bibnamefont {Zamora}},\ }\bibfield  {title} {\bibinfo {title} {{Anisotropic photon emission from gluon fusion and splitting in a strong magnetic background: The two-gluon one-photon vertex}},\ }\href {https://doi.org/10.1103/PhysRevC.106.064905} {\bibfield  {journal} {\bibinfo  {journal} {Phys. Rev. C}\ }\textbf {\bibinfo {volume} {106}},\ \bibinfo {pages} {064905} (\bibinfo {year} {2022})},\ \Eprint {https://arxiv.org/abs/2209.09364} {arXiv:2209.09364 [hep-ph]} \BibitemShut {NoStop}%
\end{thebibliography}%

\end{document}